# Cuando la ortodoxia no es lo más relevante:
# El paisaje de La Gomera y la orientación de sus iglesias [1]

A. Di Paolo (1), A. Gangui (2), J. A. Belmonte (3), M. A. Perera Betancort (4)

(1) Universidad de Buenos Aires, Facultad de Ciencias Exactas y Naturales, Departamento de Física, Buenos Aires, Argentina.
(2) Universidad de Buenos Aires, Facultad de Ciencias Exactas y Naturales, Argentina. CONICET - Universidad de Buenos Aires, Instituto de Astronomía y Física del Espacio (IAFE), Argentina.
(3) Instituto de Astrofísica de Canarias y Universidad de La Laguna, Tenerife, España.
(4) Departamento de Ciencias Históricas de la Escuela Universitaria de Turismo de Lanzarote, Universidad de Las Palmas de Gran Canaria, España.

Presentamos un estudio de la relación entre astronomía y paisaje centrado en la orientación de las iglesias y ermitas cristianas de la isla de La Gomera, situada en el Archipiélago Canario (España). El trabajo de campo consistió en la medición de las coordenadas precisas de ubicación de 38 iglesias y ermitas, lo que representa la casi totalidad de las construcciones religiosas de la isla, que cuenta con una superficie de aproximadamente 370 km². Para cada iglesia se midió, además, el acimut y la altura angular del horizonte, tomados en la dirección hacia donde apunta el altar de cada templo. Los datos así medidos fueron luego corroborados con modelos digitales de terreno de uso frecuente en estudios arqueoastronómicos. Finalmente, para el estudio de la muestra, se realizaron diversos análisis: estadístico, calendárico y orográfico, tratando de hallar indicios que permitieran comprender el patrón de orientaciones encontrado. A partir de este análisis, podemos afirmar que en algunos lugares de la isla se respetó la tradición canónica de orientar los templos cristianos en el rango solar. Asimismo, es posible que algunas pocas construcciones se orientasen con patrones de imitación del culto aborigen, especialmente en direcciones solsticiales. Sin embargo, encontramos que la orientación de la mayoría de las ermitas e iglesias es hacia el noreste y, a falta de una justificación mejor, pensamos que la razón debe buscarse más en el paisaje terrestre que en el celeste. A juzgar por la manera en la que se distribuyen varios pequeños grupos de templos -por ejemplo, aquellos ubicados en los barrancos de Hermigua y Valle Gran Rey-, estimamos que este poco habitual patrón global de orientaciones está motivado por la particular orografía de la isla. Una proporción importante de iglesias y ermitas parece adaptarse a los sitios particulares de sus emplazamientos, orientándose de acuerdo a los numerosos accidentes geográficos -en particular, sus profundos barrancos- donde están situadas. Estos resultados nos permiten conjeturar que la conocida "abrupta naturaleza" de La Gomera es quizá la principal causante del patrón particular de orientaciones de sus centros de culto.

We present a study of the relationship between astronomy and landscape centered on the orientation of Christian churches of the island of La Gomera, located in the Canary Archipelago (Spain). The fieldwork consisted of measuring the precise coordinates of 38 churches, which represents almost all of the island's religious constructions, which has an area of approximately 370 km². For each church, we measured the azimuth and the angular height of the horizon taken in the direction towards which the altar of each temple points. The data thus obtained were then corroborated with digital terrain models frequently used in archaeoastronomical studies. Finally, for the study of the sample, various analyzes were carried out: statistical, calendarical and orographic, trying to find clues that would allow us to understand the pattern of orientations found. From this analysis, we can infer that in some places on the island the canonical tradition of orienting Christian temples in the solar range was respected. Also, it is possible that a few constructions were oriented with imitation patterns of the aborigine cult, especially in solstitial directions. However, we find that the orientation of the majority of the churches is towards the northeast and, in the absence of a better justification, we think that reason should be sought more in the terrestrial landscape than in the celestial one. Judging by the way in which several small groups of temples are distributed -for example, those located in the ravines of

---

[1] El presente trabajo amplía y actualiza los resultados preliminares publicados en Di Paolo & Gangui (2018).

Hermigua and Valle Gran Rey-, we estimate that this unusual pattern of global orientations is motivated by the particular orography of the island. A significant proportion of churches seems to adapt to the specific characteristics of their sites, orienting themselves according to the numerous geographical features -their deep ravines, in particular- where they are located. These results allow us to conjecture that the known "abrupt nature" of La Gomera is perhaps the main reason for the peculiar pattern of orientations of its worship sanctuaries.

Palabras Clave: La Gomera, iglesias, arqueoastronomía, orografía.

**Introducción**

La isla de La Gomera es una de las ocho islas que, junto a cinco islotes, conforman el Archipiélago de Canarias, situado en el Océano Atlántico, en el extremo noroeste del continente africano. Se piensa que la isla fue poblada por una tribu norteafricana de origen bereber, posiblemente en torno al cambio de era, al igual que sucedió con otras seis de este archipiélago. La primera referencia que se conoce de ella es de Juba II Rey de Mauretania (25 a.C. - 23 d.C.), quien la denominó *Iunonia minor* y luego, en una fecha muy posterior, su existencia fue dibujada en el portulano de Angelino Dulcert de 1339 bajo el nombre de *Gommaria*.

La colonización de esta isla se desarrolló en diferentes etapas. Los primeros europeos llegaron a la isla en 1341 en una expedición portuguesa al mando del navegante genovés Nicolosso da Recco. A partir de ese momento marinos europeos comenzaron a visitar el archipiélago esporádicamente para proveerse de alimentos y otros bienes, tales como madera, resina y mano de obra. El investigador Álvarez Delgado (1960: 451) estima que a lo largo del siglo XIV desembarcaron en La Gomera algunos misioneros, así como diversas embarcaciones de contrabandistas y otros navegantes que solían permanecer en ella poco tiempo, y sin que sus presencias significaran conquista, colonización u ocupación, hasta el año 1420 cuando el normando Maciot de Béthencourt intentó conquistarla.

Esta última fecha es importante para el tema que nos ocupa por dos motivos:

- El inicio de la conquista de La Gomera supuso el bautizo de cada jefe de las cuatro demarcaciones indígenas de la isla, que nunca llegó a ser conquistada sino sometida. Según el historiador Aznar Vallejo (1986: 200) "el dominio de La Gomera no se fundamenta en la victoria militar, sino en la imposición de un poder superior, cimentado en el apoyo de algunas de las parcialidades de la isla y en la construcción de una torre particularmente fuerte y presta a recibir los refuerzos del exterior".

- La fecha coincide con la llegada a La Gomera de franciscanos asentados en la vecina isla de Fuerteventura, para iniciar su cristianización.

Álvarez Delgado (1960: 474), apoyado en las bulas de Benedicto XIII referidas al convento franciscano de Fuerteventura y a la evangelización emprendida en 1416 en dos islas, propone que la conversión religiosa de La Gomera comenzó entre 1417 y 1423 y que la ermita o capilla que se cita en la Bula de Martín V, del 20 de noviembre de 1424, es la de Santa Lucía, en Tazo. Sin embargo, y a pesar del escaso margen temporal, Díaz Padilla (2005: 365) estima que la obra de conversión de la población indígena pudo iniciarse con anterioridad a 1402, cuando la expedición franco-normanda llegó a la isla de Lanzarote para iniciar su conquista. Esta investigadora señala igualmente que en todo caso la población indígena conservó "muchas de sus ancestrales costumbres", tal y como lo evidenciaban los antropónimos de indígenas bautizados que añadían a su propio nombre el que recibían con el bautizo, así como topónimos o diversos testimonios recogidos en fuentes documentales.

La población indígena de La Gomera, al igual que en muchos otros territorios coloniales, se encontraba en un estado de sometimiento a la religión católica. Sin embargo, se cree que la gran mayoría de los aborígenes no vivía según los usos y costumbres propios del cristianismo. Así lo señala Díaz Padilla (2005: 366): "pero en lo relativo a la fe vivían como querían e incluso a algunos no los consideraban verdaderos cristianos". Esta particularidad fue la excepción a lo que sucedió en las demás islas del archipiélago, en las que la población aborigen desapareció o adoptó tempranamente las costumbres de los colonizadores. Algunos historiadores, como por ejemplo Aznar Vallejo (1986: 205), sostienen que tanto la abrupta orografía de La Gomera como la ausencia de una marcada conquista militar mantuvieron a la población autóctona aislada de la europea, durante un tiempo mayor que en el resto del archipiélago.

En una primera etapa de la colonización europea de La Gomera, las creencias religiosas cristianas y la cosmogonía indígena se fusionaron y se conservaron en forma de tradición oral (Navarro Mederos 2007). Los conquistadores castellanos utilizaron diversos mecanismos que facilitaron el inicio de la cristianización de la isla, como la aparición de la Virgen de Guadalupe en Punta Llana recogida en documentos escritos desde mediados del siglo XVI, y sobre la que existen otros muchos relatos en la oralidad. Este hecho presenta similitudes con la aparición de la Virgen de la Candelaria en la costa de Chimisay, en el término de Güímar en la isla de Tenerife. Igualmente, destaca la combinación de creencias aborígenes con hechos pautados por los castellanos para afianzar la cristianización insular, tal y como lo narra el ingeniero Leonardo Torriani (1978: 204) "[…] que en el cielo había un Dios llamado Orahan, quien había hecho todas las cosas; y también decía que, después de su muerte, vendrían a la isla hombres nuevos, quienes les dirían a quién debían de adorar; y decía que el hombre velludo a quien adoraban, no era el verdadero Dios de los gomeros, sino su enemigo."

A grandes rasgos, éste fue el contexto cultural en el cual se enmarcó la conquista y evangelización de La Gomera, en la que, paralelamente, se fueron edificando ermitas e iglesias católicas. El presente estudio nos permitirá analizar de qué manera este contexto pudo influenciar la construcción de estos templos religiosos, y si esto se vio plasmado en un patrón de orientaciones particular. Para ello, a continuación, caracterizaremos algunos aspectos de estos edificios de culto.

**Las ermitas y las iglesias de La Gomera**

La isla de La Gomera cuenta con un abundante patrimonio arquitectónico religioso que data del siglo XVI en adelante. Los primeros templos cristianos eran pequeñas ermitas de edificación sencilla, que fueron construidas en las distintas regiones de la isla a medida que avanzaba la colonización. Algunas fueron emplazadas dentro de los nacientes cascos urbanos, como son los casos de las ermitas San Juan Bautista, en el pueblo norteño de Vallehermoso, y San Sebastián, ubicada en la ciudad homónima. Otras fueron fundadas en lugares más periféricos, como la ermita San Isidro, en Roque Calvario (Alajeró), en la cima de la montaña de Tagaragunche (Fig. 1 izq.).

Con el correr de los siglos, a algunas de estas pequeñas ermitas se les fueron agregando capillas en la cabecera, sacristías a los lados o diversos ornamentos en la fachada. A otras se las reconstruyó completamente, alcanzando un carácter más monumental. Un ejemplo de ello es Nuestra Señora de la Asunción, en San Sebastián, que en sus inicios fue una diminuta ermita de ladrillo y mampuesto, de reducida capacidad, con disposición diferente a la actual (Fig. 1 der.). La misma hoy posee tres naves y techumbre de madera de estilo mudéjar, y fue fabricada a fines del siglo XVI (Díaz Padilla y Rodríguez Yanes 1990, Díaz Padilla 2005).

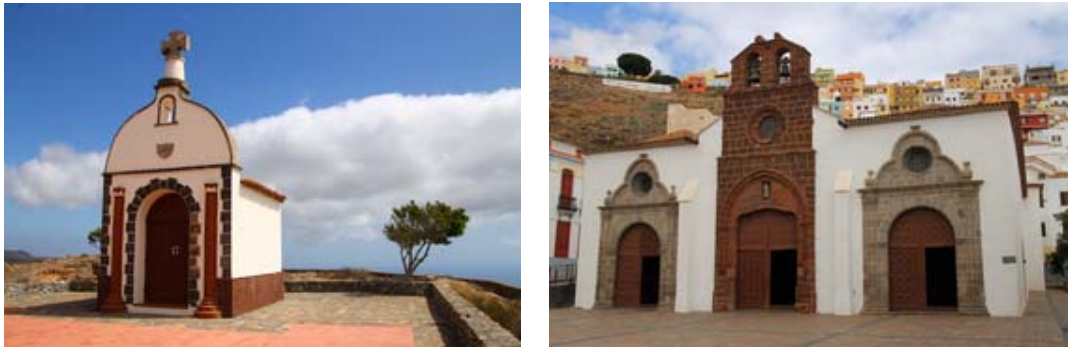

**Figura 1: Algunas iglesias características de la isla: a la izquierda San Isidro, emplazada en la cima de Roque Calvario, en el municipio de Alajeró. A la derecha Nuestra Señora de la Asunción, en San Sebastián, ciudad capital de La Gomera (Di Paolo y Gangui 2018).**

En La Gomera existen alrededor de cuarenta de estas construcciones históricas, número que resulta significativo si se tiene en cuenta que el territorio insular tiene una superficie de solamente 370 km². El estudio de la orientación astronómica de estas ermitas e iglesias, en un territorio acotado y lejos de la metrópoli, nos ofrece la oportunidad de verificar si las orientaciones típicas halladas en Europa se trasladaron rígidamente a esta colonia, o si hubo influencias de la cultura aborigen preexistente. La misma tenía notables patrones de culto (Belmonte *et al.* 1994), y una de nuestras hipótesis es que podrían haber sido conciliados con la cultura europea en un proceso de sincretismo religioso. En ese contexto, nuestro propósito fue analizar si esto se vio reflejado en algún patrón de orientaciones particular.

El trabajo de campo consistió en la medición precisa de la orientación astronómica (acimut y altura del horizonte) y de la ubicación geográfica (latitud y longitud) del conjunto de las iglesias y ermitas de la isla. Realizamos las mediciones de acimut y altura del horizonte con un tándem compuesto por brújula de precisión y clinómetro, para las cuales estimamos que los errores están en torno a 1/2º, y para la posición de cada sitio se utilizó tecnología de geolocalización. La totalidad de los datos medidos y calculados para cada iglesia y ermita se encuentran en las tablas 1 y 2, que se presentan más adelante.

En la tabla 1 se muestran los datos obtenidos en la campaña, ordenados en acimut creciente, donde cada templo cuenta con su latitud y longitud, el acimut astronómico y el nombre de la iglesia, el cual refleja el santo patrón al cual está dedicada. El acimut astronómico viene dado por la orientación de los ejes de las construcciones medida en cada sitio en dirección hacia el altar, luego corregido por declinación magnética. Los valores de la declinación magnética para los distintos sitios de la isla estaban entre 5º 27' y 5º 33' oeste en las fechas en las que se realizaron las mediciones. En la última columna de la tabla se incluye la altura angular del horizonte. En la tabla 2 se completa esta información con la declinación calculada, el día festivo del santo correspondiente, las fechas estimadas (teniendo en cuenta el año de construcción de cada iglesia) cuando la declinación del Sol es la indicada (Orientación) y la fecha en la cual fue construido cada templo. Los valores de la altura angular del horizonte fueron medidos a lo largo del eje de cada edificio en dirección hacia el altar. En aquellos sitios donde el horizonte estaba bloqueado (señalados con B en la última columna de la tabla 1), hicimos una reconstrucción del horizonte usando un modelo digital del terreno disponible en el sitio web heywhatsthat.com. El mismo posee una herramienta que, basada en el modelo de elevación proporcionado por SRTM, permite la visualización del perfil de altitud del horizonte para una ubicación dada en latitud y longitud.

Como veremos más adelante, para complementar el estudio de orientaciones astronómicas también analizamos el entorno paisajístico de cada uno de los edificios; de esta forma tuvimos en cuenta para nuestro análisis tanto el paisaje celeste como el paisaje terrestre.

| Ubicación | Nombre | L (°, N) | l(°, O) | a(°) | h(°) |
|---|---|---|---|---|---|
| (1) El Cercado | Virgen del Pino | 28.11905 | 17.28493 | 2.5 | 2.5 |
| (2) Vallehermoso | Virgen del Carmen | 28.15457 | 17.26933 | 2.5 | 13.5 |
| (3) Imada | Santa Ana | 28.08408 | 17.24084 | 2.5 | 19.0 |
| (4) Alajeró | Ntra. Sra. del Buen Paso | 28.08695 | 17.24917 | 8.5 | B 0.0 |
| (5) Las Rosas | Santa Rosa de Lima | 28.18282 | 17.22560 | 8.5 | 8.0 |
| (6) Vallehermoso | San Juan Bautista | 28.18105 | 17.26570 | 15.5 | 12.5 |
| (7) La Dama, Vallehermoso | Ntra. Sra. de las Nieves | 28.05208 | 17.30060 | 20.5 | 10.0 |
| (8) Playa de Hermigua | Sta. Catalina de Alejandría | 28.17927 | 17.18222 | 28.5 | 0.0 |
| (9) San Sebastián | Ntra. Sra. de la Inmaculada Concepción | 28.08889 | 17.11480 | 36.0 | 6.0 |
| (10) Valle Gran Rey | Ermita de los Santos Reyes | 28.10618 | 17.32376 | 40.0 | 14.5 |
| (11) Agulo | Ntra. Sra. de las Mercedes | 28.18890 | 17.19412 | 42.5 | 0.0 |
| (12) Hermigua | San Juan Bautista | 28.16167 | 17.20231 | 48.5 | B -0.5 |
| (13) Hermigua | Ntra. Sra. de la Encarnación | 28.16863 | 17.19410 | 49.5 | 0.0 |
| (14) Playa de Valle Gran Rey | San Pedro Apóstol | 28.09509 | 17.34207 | 52.0 | 20.5 |
| (15) Arure | Ntra. Sra. de la Salud | 28.13326 | 17.31979 | 53.5 | 5.0 |
| (16) San Sebastián | Ermita de San Sebastián | 28.09379 | 17.11232 | 53.5 | 26.5 |
| (17) San Sebastián | Ntra. Sra. de la Asunción | 28.09258 | 17.11121 | 59.5 | 19.5 |
| (18) Valle Gran Rey | San Antonio de Padua | 28.11781 | 17.31298 | 60.5 | 19.5 |
| (19) Las Hayas | Ntra. Sra. de Coromoto | 28.12982 | 17.29000 | 66.5 | 10.5 |
| (20) Alajeró | Santiago Apóstol | 28.03289 | 17.19202 | 66.5 | 17.5 |
| (21) Punta Llana | Ntra. Sra. de Guadalupe (ermita antigua) | 28.12657 | 17.10388 | 66.5 | B 2.5 |
| (22) Tejiade | San José | 28.07216 | 17.19218 | 78.5 | 1.0 |
| (23) Chipude | Ntra. Sra. de la Candelaria | 28.10984 | 17.28239 | 84.0 | 7.0 |
| (24) Alajeró | El Salvador | 28.06347 | 17.24038 | 88.5 | 5.0 |
| (25) Roque Calvario, Alajeró | San Isidro | 28.05301 | 17.24199 | 109.5 | -1.0 |
| (26) Alojera, Vallehermoso | Ntra. Sra. de la Inmaculada Concepción | 28.16053 | 17.32483 | 117.5 | 20.0 |
| (27) Erque, Alajeró | San Lorenzo | 28.08335 | 17.26091 | 126.0 | 12.5 |
| (28) Benchijigua | San Juan Bautista | 28.09157 | 17.21880 | 177.5 | 16.5 |
| (29) Igualero | San Francisco | 28.09940 | 17.25491 | 231.0 | B -1.5 |
| (30) Agulo | San Marcos Evangelista | 28.19772 | 17.19783 | 239.5 | 13.5 |
| (31) Las Nieves | Ntra. Sra. de la Salud | 28.10112 | 17.20201 | 255.5 | 1.5 |
| (32) Valle Gran Rey | Ntra. Sra. del Buen Viaje | 28.13870 | 17.33758 | 285.0 | 0.0 |
| (33) Playa de Santiago | Virgen del Carmen | 28.02728 | 17.19864 | 289.5 | B 0.0 |
| (34) La Palmita | San Isidro | 28.17124 | 17.21595 | 290.0 | 13.5 |
| (35) Guarimiar | Sagrado Corazón de Jesús | 28.06678 | 17.22075 | 297.0 | 23.5 |
| (36) Parque Nac. de Garajonay | Ntra. Sra. de Lourdes | 28.12696 | 17.22085 | 302.5 | B 0.0 |
| (37) Valle Gran Rey | Virgen del Carmen | 28.08205 | 17.33348 | 309.0 | 0.0 |
| (38) Hermigua | Ntra. Sra. del Rosario (conv. San Pedro) | 28.15278 | 17.19759 | 325.5 | 13.5 |
| (39) Punta Llana | Ntra. Sra. de Guadalupe (iglesia actual) | 28.12657 | 17.10388 | 334.5 | 4.0 |

**Tabla 1: Orientaciones de las iglesias de La Gomera, ordenadas por acimut creciente. Para cada construcción, la tabla muestra la ubicación, la identificación, la latitud y longitud geográficas (L y l), el acimut astronómico (a) tomado a lo largo del eje del edificio en dirección al altar y la altura angular del horizonte (h) en la dirección del acimut correspondiente (ambos valores aproximados al 0.5° de error). Señalamos con B los valores de h reconstruidos utilizando un modelo digital del terreno. Todos los valores de la tabla están expresados en grados (Di Paolo y Gangui 2018).**

| Ubicación | Nombre | δ(°) | Fecha (santo patrón) | Orientación | Fecha construcción |
|---|---|---|---|---|---|
| (1) El Cercado | Virgen del Pino | 64.0 | 1 sep. | | 1970 |
| (2) Vallehermoso | Virgen del Carmen | 75.1 | 16 jul. | | 1697 |
| (3) Imada | Santa Ana | 80.6 | 26 jul. | | 1535 |
| (4) Alajeró | Ntra. Sra. del Buen Paso | 60.2 | 2 feb./15 sep. | | 1544 |
| (5) Las Rosas | Santa Rosa de Lima | 68.2 | 30 ago. | | 1967 |
| (6) Vallehermoso | San Juan Bautista | 68.6 | 24 jun. | | 1632 |
| (7) La Dama, Vallehermoso | Ntra. Sra. de las Nieves | 63.6 | 5 ago. | | 1954 |
| (8) Playa de Hermigua | Sta. Catalina de Alejandría | 50.3 | 1er lunes oct. | | 1735 |
| (9) San Sebastián | Ntra. Sra. de la Inmaculada Concepción | 49.3 | 8 dic. | | principios s. XVI |
| (10) Valle Gran Rey | Ermita de los Santos Reyes | 50.4 | 6 ene. | | 1533 |
| (11) Agulo | Ntra. Sra. de las Mercedes | 40.1 | 24 sep. | | 1611 |
| (12) Hermigua | San Juan Bautista | 35.4 | 24 jun. | | 1943 |
| (13) Hermigua | Ntra. Sra. de la Encarnación | 34.6 | 8 dic. | | 1911 |
| (14) Playa de Valle Gran Rey | San Pedro Apóstol | 42.4 | 29 jun. | | 1982 |
| (15) Arure | Ntra. Sra. de la Salud | 34.3 | 15 jul. | | 1965 |
| (16) San Sebastián | Ermita de San Sebastián | 42.8 | 20 ene. | | 1535-1540 |
| (17) San Sebastián | Ntra. Sra. de la Asunción | 35.4 | 15 ago. | | fin s. XVI |
| (18) Valle Gran Rey | San Antonio de Padua | 34.5 | 13 jun. | | 1992 |
| (19) Las Hayas | Ntra. Sra. de Coromoto | 25.6 | 1er quincena sep. | | 1985 |
| (20) Alajeró | Santiago Apóstol | 28.5 | 25 jul. | | 1939 |
| (21) Punta Llana | Ntra. Sra. de Guadalupe (ermita antigua) | 21.7 | 6 sep. | 20 may. ± 4 d. / 5 jul. ± 4 d. | 1542 |
| (22) Tejiade | San José | 10.5 | 19 mar. | 17 abr. ± 2 d. / 26 ago. ± 2 d. | 1719 |
| (23) Chipude | Ntra. Sra. de la Candelaria | 8.5 | 2 feb./15 ago. | 2 abr. ± 2 d. / 22 ago. ± 2 d. | 1530/40 |
| (24) Alajeró | El Salvador | 3.6 | 5 ago. | 4 sep. ± 2 d. / 20 mar. ± 2 d. | 1666/81 |
| (25) Roque Calvario, Alajeró | San Isidro | -18.0 | 15 may. | 28 ene. ± 3 d. / 13 nov. ± 3 d. | (*) |
| (26) Alojera, Vallehermoso | Ntra. Sra. de la Inmaculada Concepción | -12.9 | 8 dic. | 15 feb. ± 2 d. / 27 oct. ± 2 d. | principios s. XVII |
| (27) Erque, Alajeró | San Lorenzo | -23.9 | 10 ago. | 13 dic. ± 9 d. (a) | 1512 |
| (28) Benchijigua | San Juan Bautista | -45.4 | 24 jun. | | 1603 |
| (29) Igualero | San Francisco | -35.1 | 4 oct. | | 1984 |
| (30) Agulo | San Marcos Evangelista | -19.0 | 25 abr. | 25 ene. ± 3 d. / 17 nov. ± 3 d. | 1607 |
| (31) Las Nieves | Ntra. Sra. de la Salud | -12.1 | 7,8 sep./2do domingo oct. | 7 feb. ± 3 d. / 15 oct. ± 3 d. | mediados s. XVI |
| (32) Valle Gran Rey | Ntra. Sra. del Buen Viaje | 13.1 | último domingo ago. | 24 abr. ± 2 d. / 18 ago. ± 2 d. | 1980 |
| (33) Playa de Santiago | Virgen del Carmen | 16.9 | 16 jul. | 7 may. ± 3 d. / 5 ago. ± 3 d. | (*) |
| (34) La Palmita | San Isidro | 23.8 | 15 may. | 21 jun. ± 8 d. | 1950 |
| (35) Guarimiar | Sagrado Corazón de Jesús | 33.7 | 3 o 23 de jun. | | 1991 |
| (36) Parque Nac. de Garajonay | Ntra. Sra. de Lourdes | 28.0 | último domingo ago. | | 1935 |
| (37) Valle Gran Rey | Virgen del Carmen | 33.4 | 16 de julio | | 1987 |
| (38) Hermigua | Ntra. Sra. del Rosario (conv. San Pedro) | 54.7 | 1er dom. oct./8 ago./22 may. | | 1598/1611 |
| (39) Punta Llana | Ntra. Sra. de Guadalupe (iglesia actual) | 55.7 | 6 sep. | | 1542 |

**Tabla 2: Declinaciones de las iglesias de La Gomera. Para cada construcción, se muestra la ubicación, la identificación, la declinación astronómica resultante (δ), la fecha del santo patrón o festividad, las fechas estimadas (teniendo en cuenta el año de construcción de cada iglesia) cuando la declinación del Sol es la indicada (Orientación) y la fecha aproximada en la que fue construida. En el cuerpo de la tabla, (*) indica que, al presente, desconocemos la fecha de construcción de estas iglesias, y (a) señala que el solsticio corresponde a esta fecha del calendario de la época de construcción de la iglesia, previo a la instauración del calendario gregoriano en 1582.**

**La orientación de iglesias y el paisaje**

En la figura 2 se muestra el diagrama de orientación para las iglesias y ermitas estudiadas. Como hemos mencionado, los valores de los acimuts son los medidos, e incluyen la corrección por declinación magnética.

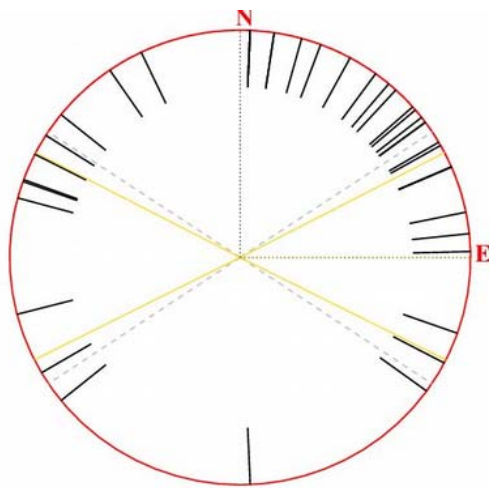

**Figura 2: Diagrama de orientación para las iglesias y ermitas de La Gomera, obtenido a partir de los datos de la tabla 1. Las líneas amarillas (líneas continuas) señalan los acimuts correspondientes -en el cuadrante oriental- a los valores extremos para el Sol (acimuts de 62,8º y 116,5º, equivalente a los solsticios de verano e invierno boreales, respectivamente) y para la Luna (acimuts: 56,7º y 123,5º -líneas rayadas-, equivalente a la posición de los lunasticios mayores). Aunque varias construcciones -unas doce- siguen la orientación canónica en el rango solar, un gran número de iglesias se alinean al noreste (Di Paolo y Gangui 2018).**

De las 39 orientaciones medidas en las ermitas e iglesias, 13 se dirigen hacia el cuadrante norte, 9 hacia el cuadrante occidental, 16 hacia el oriental y tan solo una hacia el cuadrante meridional. De todas estas, 12 orientaciones se ubican en el rango solar, ya sea a levante (7) o a poniente (5), que representan menos de un tercio del total. Las 7 iglesias orientadas *ad orientem* tienen la particularidad de ser iglesias de tipología colonial. Algunas de ellas son, por ejemplo, la ermita Nuestra Señora de Guadalupe, en Punta Llana, o Nuestra Señora de la Candelaria, en Chipude, ambas del siglo XVI.

Del diagrama de orientación podemos distinguir un patrón de orientaciones claro, hacia el noreste, que no guarda precedente en otros estudios de iglesias antiguas o coloniales, a excepción quizá de aquellas de Lanzarote, también en el archipiélago canario. En aquella isla también se descubrió que una notable proporción de los templos estaba orientada aproximadamente hacia el norte-noreste (con "entrada" a sotavento) para evitar los vientos dominantes del lugar, los alisios, provenientes justamente de esa dirección (Fig. 3 izq.).[2] Como una muestra de este fenómeno climático, en la figura 3 (der.) se puede observar el diagrama de vientos tomado del aeropuerto de Arrecife de Lanzarote (Gangui *et al*. 2016a). Pero en el presente caso de estudio, en La Gomera, ese factor "práctico" lanzaroteño, debido a la particular combinación de geografía y clima, no es preponderante.

---

[2] Lanzarote posee una amplia región en la zona central llamada El Jable, que está caracterizada por poseer un suelo de tipo arenoso. Se piensa que la presencia de esta amplia región con abundancia de arenas, sumado a los fuertes vientos procedentes de dirección noreste, motivó la construcción de las edificaciones cercanas a El Jable con la puerta orientada en dirección opuesta al viento, es decir, con el altar apuntando hacia el noreste.

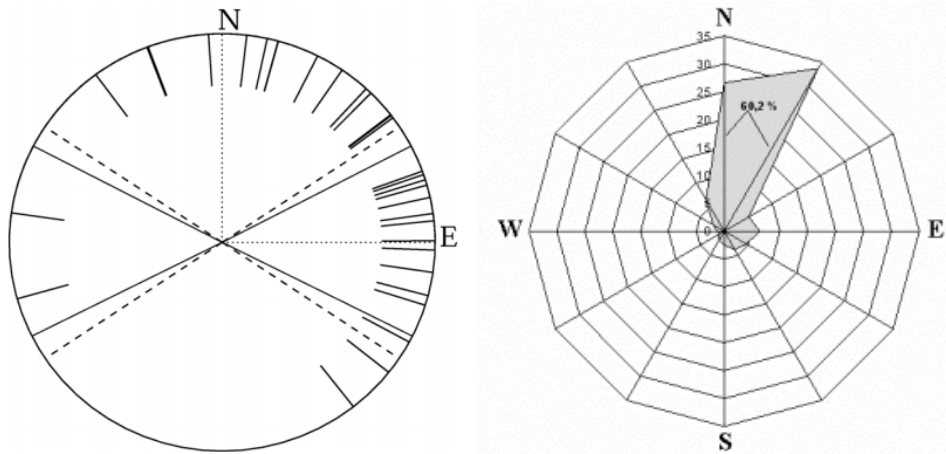

**Figura 3:** Diagrama de orientación para las iglesias de Lanzarote (izquierda) y diagrama de vientos característico para esa misma isla (derecha), ilustrativo de los vientos alisios dominantes que se piensa han llevado a su población a construir las iglesias apartándose de las orientaciones "canónicas" (Gangui *et al.* 2016a).

Como mencionamos anteriormente, las orientaciones de las iglesias gomeras tienen poco correlato con otros estudios previos. En la Península Ibérica y en todo el Mediterráneo, estas construcciones se orientaban mayoritariamente en el rango solar (González-García 2015), tal como lo prescribían escritores y apologetas del cristianismo temprano. Como detallaremos más adelante, estimamos que la particular orografía de La Gomera fue la que marcó este peculiar patrón de orientaciones, alejado de los patrones canónicos.

**Análisis estadístico**

Para poder abordar en detalle posibles orientaciones astronómicas, debemos emplear las declinaciones en lugar de los acimuts. Por ello, en la figura 4 presentamos un histograma de las declinaciones cuyos valores pueden calcularse a partir de los datos de la tabla 1. Este análisis estadístico nos permite estudiar el patrón de orientaciones de manera más precisa, ya que la declinación astronómica es independiente de la ubicación geográfica y de la topografía local. El gráfico muestra la declinación astronómica frente a la frecuencia relativa normalizada, lo que nos permite una más clara y acertada determinación de la estructura de picos y la relevancia estadística de los mismos. Los valores de declinación correspondientes a cada iglesia con los que está elaborado el gráfico, se pueden observar en la tabla 2. Para nuestro caso de estudio tenemos tres hipótesis fundamentales sobre la posible orientación de las iglesias.

La primera hipótesis indica que los templos están orientados con su altar apuntando hacia el este geográfico, es decir, hacia la salida del Sol en los equinoccios. De cumplirse esta hipótesis se observaría en el gráfico un pico (> 3σ) en torno a una declinación aproximada de 0º.

La segunda hipótesis que debemos considerar es la que sugiere que las iglesias se orientan hacia la salida del Sol en el día de los solsticios. Esta orientación puede ser considerada un patrón de imitación del culto aborigen prehispánico (Belmonte *et al.* 1994). De cumplirse esta hipótesis, se vería reflejada como una notoria acumulación de orientaciones en torno a una declinación de ±23,5º.

La última hipótesis es la que propone que las iglesias y ermitas están orientadas hacia la salida del Sol en el día correspondiente a su santo patrón o festividad. En ella las declinaciones resultantes toman valores arbitrarios dentro del rango (-23,5º, +23,5º) y, por lo tanto, no se vería ningún pico relevante en el gráfico (a menos que varias iglesias tuvieran la misma fiesta patronal).

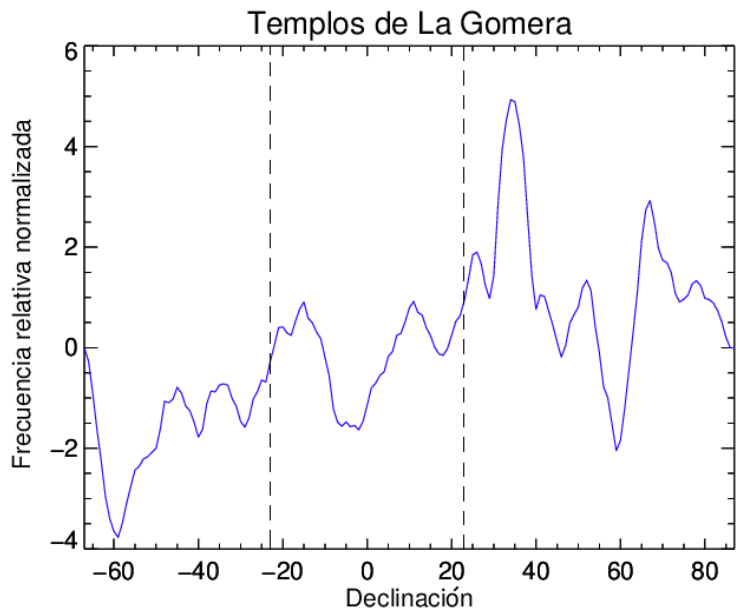

**Figura 4: Frecuencia relativa normalizada de los valores de la declinación astronómica para las ermitas e iglesias de La Gomera. Se aprecian dos picos con amplitudes relevantes, uno bien por encima de 4σ, en δ ~ 35º ± 1º, y otro un poco menor a 3σ, en δ ~ 66º ± 1º.**

Como se puede apreciar, el gráfico presenta un pico principal (cercano a 5σ) en torno a los 35º ± 1º, declinación que se encuentra fuera del rango solar.[3] Como es superior a 3σ reviste importancia estadística, pero, por no ser una orientación hacia la salida del Sol en alguna fecha, estadísticamente no se encuentra una explicación dentro de las hipótesis mencionadas. Algunas de las iglesias que poseen orientaciones con declinación cercana al valor 35º ± 1º son: Nuestra Señora de la Asunción en San Sebastián (número 17 en la tabla 2) y una más moderna, Nuestra Señora de la Encarnación en Hermigua (núm. 13).

Por otro lado, existe un segundo pico de menor relevancia que el anterior que llega hasta un valor cercano a 3σ en torno a los 66º ± 1º. Esta declinación corresponde a una orientación ubicada más al norte y, por lo tanto, también fuera del rango solar. Esta amplitud podría asociarse a un pico de acumulación por orientaciones cercanas a la línea meridiana. Algunas iglesias orientadas con esta declinación son: Santa Rosa de Lima, en Las Rosas (núm. 5), y San Juan Bautista, en Vallehermoso (núm. 6).

De esa forma, el estudio estadístico de declinaciones confirma lo observado en el diagrama de orientación de la figura 2, es decir, que existió una aparente predilección para orientar las iglesias hacia el noreste. Por otro lado, para los templos que se encuentran dentro del rango solar, se realizó un análisis calendárico individual, como se detalla a continuación.

**Análisis calendárico**

---

[3] El error consignado para los valores de declinación surge de la propagación de errores en el cálculo de la medición indirecta (declinación), a partir de las mediciones directas (acimut y altura del horizonte). Aunque estos errores son diferentes para cada iglesia, tomamos el error promedio (1º), para nuestro análisis estadístico.

Una de las teorías subyacentes a la orientación de ermitas e iglesias, sugiere que estas tienden a construirse con el altar apuntando en la dirección de la salida del Sol en el día de su santo patrón o festividad. Mediante un cálculo de efemérides, intentamos corroborar está hipótesis para las iglesias de la isla.

La tabla 2 muestra los valores de declinaciones calculados para la totalidad de las iglesias. En ella, tenemos 12 iglesias orientadas dentro del rango solar, es decir, cuya declinación calculada se halla aproximadamente entre +24º y -24º (teniendo en cuenta en este rango las que se encuentran dentro del error en declinación). En la columna *Orientación* de la tabla 2, se observan las fechas correspondientes a cada declinación solar, en el año en el cual fue construido cada templo.[4] El error calendárico (en días) que consignamos para cada iglesia, proviene de la estimación de la cantidad de días que tarda el Sol en desplazarse el correspondiente error en declinación. Un ejemplo de esto es el caso de Nuestra Señora de la Salud en Las Nieves (núm. 31), donde se obtuvo una declinación de -12,1º, con un error calculado de aproximadamente 0.6º. El paso solar por esta declinación fue, a mediados del s. XVI (época en la que fue construida la iglesia), en los días 7 de febrero ±3 días o 15 de octubre ±3 días. A su vez, la advocación a la Virgen de La Salud se celebra el segundo domingo de octubre. Por lo tanto, podría tratarse de un caso en el que hubo una intencionalidad en cuanto a la orientación de la iglesia según su fecha festiva. Sin embargo, por tratarse de un sólo caso, no se puede descartar que sea una coincidencia y, además, el segundo domingo de octubre no es una fecha precisa porque, por supuesto, cambia de día año tras año. Un caso peculiar es el de la ermita San Lorenzo en Erque (Alajeró) (núm. 27) cuya orientación es solsticial, ya que el día en el que se ubica el Sol en la declinación calculada (tomando en cuenta el rango de error) es el 13 de diciembre ±9 días. La otra iglesia solsticial de nuestra tabla de datos es la de San Isidro, en La Palmita (núm. 34), pero en este caso la construcción es mucho más reciente. En el resto de las iglesias que se encuentran dentro del rango solar no se cumple la hipótesis que señala correlación entre orientación y fiesta patronal. Así, con el análisis calendárico y estadístico del conjunto estudiado, descartamos que haya existido una tradición sostenida en el tiempo (salvando los casos mencionados) de orientar las iglesias hacia la salida del Sol en fechas específicas.

Sin embargo, para caracterizar el particular patrón de orientaciones encontrado tanto en acimut como en declinaciones (con una gran frecuencia de orientaciones fuera del rango solar), es preciso complementar este estudio con esfuerzos dedicados a ponderar la posible influencia del paisaje y la orografía circundante a los templos. Esto será tratado a continuación en un análisis orográfico de la isla (cf. Di Paolo y Gangui 2018).

**Análisis orográfico**

Para visualizar mejor la posible influencia sobre las orientaciones de la orografía, en forma de accidentes naturales como valles y barrancos, o incluso por la presencia de algún monte prominente de la isla (como el Roque de Imada)[5] o de islas cercanas (como el volcán Teide de la isla de Tenerife, bien visible desde La Gomera), en la figura 5 presentamos un mapa topográfico que incorpora las orientaciones de todas las iglesias estudiadas.[6] En el mismo se aprecian grupos de iglesias que "copian" la dirección de los valles profundos en los que se hallan emplazadas. Este es el caso de la iglesia de Nuestra Señora de la Encarnación (número 13 en la tabla 1), y de las ermitas de San Juan Bautista y Santa Catalina (núm. 12 y 8), todas ellas ubicadas a lo largo del barranco (o valle) de

---

[4] Para los cálculos de efemérides se utilizó el software XEphem 3.7.7. donde, además, tuvimos en cuenta el cambio al calendario juliano que entró en vigencia en 1582 en España y sus colonias.
[5] La palabra "roque", empleada en Canarias, denota una elevación rocosa y muy escarpada.
[6] El mapa fue realizado a partir del procesamiento de imágenes LANDSAT, cortesía de USGS (Servicio Geológico de los Estados Unidos).

Hermigua. Lo mismo sucede con los templos de San Antonio de Padua (núm. 18), San Pedro Apóstol (núm. 14) y el de los Santos Reyes (núm. 10), también ubicados a lo largo de un barranco prominente, pero que esta vez se trata del Valle Gran Rey, que vierte en la costa oeste de la isla. En estos dos casos los barrancos siguen una línea sudoeste-noreste, coincidente con la acumulación de orientaciones en la región noreste del diagrama de la figura 2. A esto se suma la presencia en la propia capital de la isla, San Sebastián de La Gomera, de tres construcciones con orientaciones en el cuadrante nororiental: las ermitas de Nuestra Señora de la Inmaculada Concepción (núm. 9) y de San Sebastián (núm. 16), y la iglesia matriz de Nuestra Señora de la Asunción (núm. 17) que, con buena aproximación, refuerzan ese patrón. Cabe mencionar que estos últimos templos fueron edificados según el planeamiento original de la ciudad y que, posteriormente, a pesar de que la ciudad fue invadida y destruida en varias oportunidades, su trazado original se mantuvo con el transcurrir de los siglos. Así lo afirma el historiador gomero Alberto Darías Príncipe: "(...) las mismas manzanas, las mismas huertas, las mismas calles y los mismos espacios libres. Todo se reconstruye donde estaba antes del desastre, pero se tiende a completar y complementar la antigua trama urbana." (Darías Príncipe 1992).

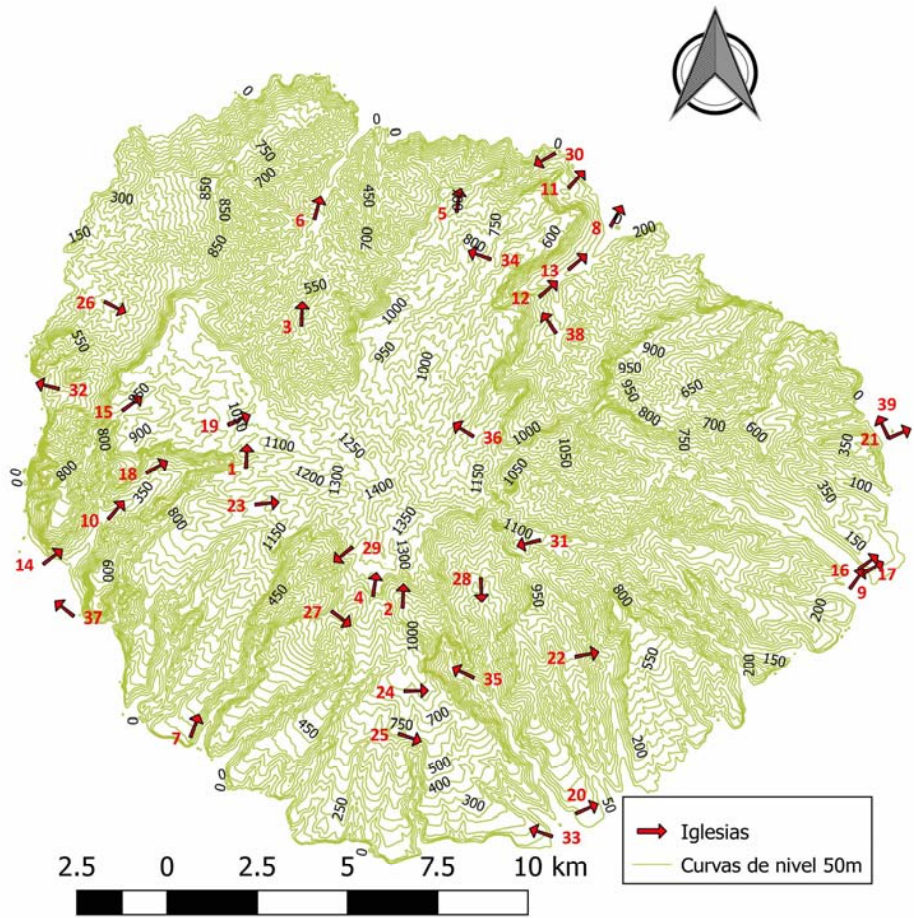

**Figura 5: Mapa topográfico con la ubicación geográfica de la totalidad de las iglesias medidas, junto con la orientación de sus ejes en dirección al altar (flechas, orientadas de acuerdo a los acimuts consignados en la tabla 1). En la región de Punta Llana (costa este) la iglesia Nuestra Señora de Guadalupe presenta dos flechas con orientaciones ortogonales. Estas se deben a que, adosada a la antigua ermita (núm. 21), que inicialmente apuntaba con acimut 66,5º en dirección al volcán Teide, en la vecina isla de Tenerife, tiempo más tarde se construyó la nave principal de**

**la iglesia actual (núm. 39), con una orientación aproximadamente perpendicular (Di Paolo y Gangui 2018).**

La orografía también permite comprender la disposición de otras construcciones que podríamos llamar "anómalas", en el sentido de que se apartan mucho de las esperadas, si nos atenemos a la tradición religiosa canónica de orientaciones. Ejemplo de esto es la iglesia de Santa Ana, en Imada, que se orienta prácticamente con la meridiana del lugar (con un acimut de 2,5º, en dirección norte). Una inspección del paisaje circundante, sin embargo, muestra que el eje de la iglesia se alinea con muy buena aproximación con la elevación montañosa más distintiva del lugar: el Roque de Imada. Pero en este caso, el roque se halla del lado de la puerta de la iglesia, en dirección opuesta a la del altar, por lo que su visión se vuelve imponente al salir del edificio, luego de culminar el oficio religioso. Guardando las distancias -tanto geográficas como por tratarse de diferentes regiones culturales-, casos de contribuciones similares a estos ya fueron descritos y estudiados con cierto detalle en el norte de Chile (Gangui, Guillén y Pereira 2016b).

Otro ejemplo es el de la ermita de Nuestra Señora del Buen Paso, en Alajeró, cuyo eje tiene un acimut de 8,5º y, por lo tanto, se aparta mucho de las orientaciones canónicas. Nuevamente, en este caso, la inspección del paisaje circundante nos indica que el movimiento del Sol, y el arco de sitios del horizonte por donde surge o se oculta en diferentes días del año, fue irrelevante para quienes concibieron la iglesia; pues la construcción se ubica pegada a la ladera de un barranco escarpado, con el fin presumible de indicar el mejor camino para atravesar la montaña, y por supuesto no hubo libertad para orientarla adecuadamente.

**Discusión**

Los templos cristianos de La Gomera empezaron a edificarse a partir del sometimiento y colonización de la isla. Este proceso comenzó, en una primera fase, a principios del siglo XV, y fue llevado a cabo por miembros de la nobleza católica. Posteriormente, hacia fines del mismo siglo, fue continuado por la ocupación realenga, impulsada y financiada directamente por los Reyes Católicos de la Corona de Castilla. Ambas etapas llevaron la religión católica a la isla y la evangelización de la población nativa, con la consiguiente construcción de pequeñas ermitas y templos que ilustrarían una nueva situación religiosa y social. Estos sucesos se dieron como un paso previo a la conquista del continente americano, lo que resulta interesante porque los patrones encontrados para la orientación de iglesias en esta isla podrían ser una "muestra" de lo que sucedería posteriormente durante la colonización de América (Macías Hernández 1994).

Como antecedente sobre el estudio de templos cristianos antiguos, se sabe que la orientación espacial de los mismos es una de las características más destacadas de su arquitectura. Las prescripciones, dadas en textos del cristianismo temprano, indicaban que las iglesias debían situarse siguiendo una determinada orientación, con el sacerdote mirando hacia oriente durante los oficios. Trabajos recientes muestran que estas prescripciones se cumplieron, al menos durante el Medioevo, donde hubo una marcada tendencia a orientar los altares de los templos en el rango solar. Este es el contexto de nuestro presente estudio arqueoastronómico, y nuestra intención fue investigar si las mencionadas tradiciones fueron trasladadas rígidamente desde el continente europeo hacia las nuevas colonias o si, por el contrario, estas prescripciones no se respetaron y por qué razón.

Comenzamos con un estudio de campo en el cual relevamos un total de 38 ermitas e iglesias distribuidas en una superficie de 370 km². y esto fue complementado con análisis específicos detallados. El estudio de las orientaciones acimutales arrojó que de 39 orientaciones medidas (38 construcciones, pero donde una de ellas –Nuestra Señora de Guadalupe en Punta Llana– mantiene como sacristía una antigua ermita con orientación perpendicular a la de la iglesia principal), 12 se encuentran en el rango solar (7 hacia el levante y 5 hacia el poniente). Cabe destacar que de estas doce iglesias la mayoría son de épocas anteriores al siglo XVIII, con las excepciones de Nuestra Señora del Buen Viaje en el Valle Gran Rey y de San Isidro en la Palmita, que son del siglo XX.

Por otro lado, hemos podido comprobar que la mayoría de los templos se orienta con sus ejes apuntando hacia el cuadrante noreste. A partir de nuestro análisis deducimos que esta poco habitual orientación en los templos muy probablemente estuvo relacionada con factores prácticos que prevalecieron sobre las tradiciones, como mencionaremos más adelante. El análisis estadístico de las declinaciones astronómicas presentado aquí es más apropiado metodológicamente que el solo uso de las orientaciones acimutales, pues tiene en cuenta también los valores de las alturas del horizonte local presentados en la tabla 1. Este análisis dio resultados que muestran una mayor frecuencia de los ejes de los templos orientados hacia declinaciones cercanas a los 35º, hacia el hemisferio norte celeste y fuera del rango solar. Además de este análisis estadístico, que amplía y corrige algunas de las hipótesis planteadas en un trabajo preliminar (Di Paolo y Gangui 2018) sobre la orientacion acimutal de las iglesias, el cálculo de las declinaciones astronómicas nos permitió realizar un análisis calendárico individual para cada iglesia orientada dentro del rango solar. Para ello, además, se requirió un estudio de los registros históricos para precisar el año de fundación de cada templo. A partir de estos análisis hemos podido comprobar que la mayoría de las iglesias de La Gomera no se hallan orientadas hacia la salida del Sol en las efemérides del santo patrón o fecha festiva particular de cada iglesia. Sin embargo, en el caso particular de la iglesia Nuestra Señora de la Salud en Las Nieves, el edificio sí podría haber sido construido con su altar apuntando hacia la salida del Sol en su día festivo. No obstante, al ser su celebración el segundo domingo de octubre, esto no permite un cálculo exacto ya que ese segundo domingo cambia de día año tras año, y la fecha de construcción de la iglesia, de mediados del siglo XVII, no se conoce con exactitud.

No se pudo establecer una correlación entre las orientaciones de las iglesias de la isla y el culto previo a la conquista. Una de las hipótesis era que, de haberse encontrado una cierta cantidad de iglesias con sus ejes alineados hacia los solsticios, habríamos podido inferir que quizás existió un proceso de sincretismo religioso (Belmonte *et al.* 1994). Esto lo sabemos por las recomendaciones que daban distintos autores cristianos en el Medioevo sobre la orientación de las iglesias, para evitar su construcción en direcciones próximas a los solsticios. Posiblemente esto ocurría por los numerosos templos paganos que existían orientados en esta dirección. Por otro lado, se sabe que los aborígenes de las Canarias tenían un patrón de culto en el cual el solsticio era importante, posiblemente influenciados por las culturas del norte de África. En nuestra investigación encontramos solamente una iglesia orientada hacia un solsticio: San Lorenzo en Erque, Alajeró. Sin embargo, al tratarse de un único caso, no podemos descartar que esto se haya dado de forma accidental, ya que el error "calendárico" (es decir, la incerteza en la determinación de una fecha en el calendario a partir de un dado valor de declinación del Sol) es comparativamente grande durante los solsticios. Asimismo, cabe remarcar que la inclusión de este análisis de declinaciones astronómicas nos permitió descartar la hipótesis planteada en un trabajo preliminar (Di Paolo y Gangui 2018), sobre una posible orientación solsticial de la iglesia de Nuestra Señora de la Inmaculada Concepción, de Alojera.

Finalmente, pudimos poner en diálogo estos nuevos aportes con el análisis orográfico ya realizado en (Di Paolo y Gangui 2018). A partir de esto y de lo mencionado en los párrafos anteriores, podemos conjeturar que muchas de las primeras ermitas e iglesias de la isla comenzaron orientándose en el rango solar (aunque esto no fue una regla), pero que con el correr de los siglos, en las iglesias más modernas, lo que prevaleció fue la adaptación a la orografía de sus sitios de emplazamiento. Creemos que la "abrupta naturaleza" de La Gomera fue la responsable de este comportamiento (Díaz Padilla 2005). Esto se puede observar con claridad en los barrancos de Hermigua y del Valle Gran Rey, donde la orientación de las iglesias "sigue" aproximadamente a la dirección de las curvas de nivel, en sentido sudoeste-noreste. Estimamos que esto podría haber tenido un sentido práctico, ya que las iglesias se imponen visualmente de mejor forma al tener su fachada apuntando en dirección de los barrancos. De esta forma, tanto los habitantes de las poblaciones situadas en los valles como los nuevos colonos podían apreciar de mejor forma a estas estructuras inmersas en el paisaje.

Las iglesias ubicadas en estos valles contribuyen a incrementar la prevalencia de las orientaciones en la dirección noreste, y a esto se suman las iglesias presentes en la ciudad de San Sebastián de La Gomera que siguieron el trazado de la red urbana de la, en aquel momento, naciente ciudad. Con respecto a la posible influencia de los vientos, hallamos (cf. Di Paolo y Gangui 2018) que no serían tan condicionantes para la orientación de las iglesias como sí lo fueron en la isla de Lanzarote (Gangui *et al.* 2016a). Sobre este punto, podemos mencionar que en las iglesias de La Gomera no se observan muros adicionales para proteger a los edificios del viento, por ejemplo, barbacanas, como sí se encuentran en la isla de Lanzarote, donde la región de El Jable condiciona construcciones y pueblos enteros. En muchos pueblos de La Gomera es la propia orografía montañosa de la isla la que protege a las construcciones de los vientos alisios provenientes del norte.

**Agradecimientos**

**Referencias**

Álvarez Delgado J.
    1960 "Primera conquista y cristianización de Las Islas Canarias. Algunos problemas históricos". *Anuario de Estudios Atlánticos*, Madrid, Las Palmas, VI: 445- 492.

Aznar Vallejo E.
    1986 "La colonización de las Islas Canarias en el siglo XV". *En la España Medieval*, T. V. En memoria de Claudio Sánchez Albornoz. Editorial de la Universidad Complutense de Madrid, I: 195-218.

Belmonte J. A., Esteban C., Aparicio A., Tejera Gaspar A. y González O.
    1994 "Canarian Astronomy before the conquest: the pre-hispanic calendar". *Rev. Acad. Can. Ciencias*, VI (2-3-4): 133-156.

Darías Príncipe A.
    1992 *La Gomera: espacio, tiempo y forma*. Compañía Mercantil Hispano-Noruega, Santa Cruz de Tenerife.

Díaz Padilla G.
    2005 "La evolución parroquial de La Gomera y el patrimonio documental generado por la institución eclesiástica". *Memoria ecclesiae*, 27: 365-376.

Díaz Padilla G. y Rodríguez Yanes J.M.
    1990 *El Señorío en las Canarias occidentales: La Gomera y El Hierro hasta 1700*. Cabildo Insular de la Gomera, San Sebastián de La Gomera.

Di Paolo A. y Gangui A.

    2018 "Estudio arqueoastronómico de las iglesias históricas de La Gomera". *Anales de la Asociación Física Argentina*, 29 (3): 62-68.

Gangui A., González-García A. C., Perera Betancort M. A., y Belmonte J. A.

    2016a "La orientación como una seña de identidad cultural: las iglesias históricas de Lanzarote". *Tabona.* Revista de Prehistoria y Arqueología, 20: 105-128.

Gangui A., Guillén A. y Pereira M.

    2016b "La orientación de las iglesias andinas de la región de Arica y Parinacota, Chile: una aproximación arqueoastronómica". *Arqueología y Sociedad*, 32: 303-322.

González-García A. C.

    2015 "A Voyage of Christian Medieval Astronomy: Symbolic, ritual and political orientation of churches". *En Stars and Stones: Voyages in archaeoastronomy and cultural astronomy.* British Archaeology Reports, Int. Ser. 2720, edited by F. Pimenta *et al.*, 268-275.

Macías Hernández A.

    1994 "Relaciones entre América y Canarias". *Gran Enciclopedia Canaria.* Santa Cruz de Tenerife, Ediciones Canarias, 1: 224-228.

Navarro Mederos, J. F.

    2007 "Santuarios y espacios sacralizados entre los antiguos gomeros". *Veleia*, 24-25.

Torriani L.

    1978 "Descripción de las Islas Canarias". Gota Ediciones, Santa Cruz de Tenerife.